\documentclass[nohyperref]{article}
\usepackage[margin=1in]{geometry}
\usepackage{authblk}
\usepackage{microtype}
\usepackage{graphicx}
\usepackage{adjustbox}
\usepackage{subfigure}
\usepackage{booktabs} 
\usepackage{wrapfig}
\usepackage{hyperref}
\usepackage{soul} 
\usepackage{multirow}

\usepackage[dvipsnames, table]{xcolor}

\usepackage{amsmath}
\usepackage{amssymb}
\usepackage{mathtools}
\usepackage{amsthm}
\usepackage{amsfonts}
\usepackage[capitalize,noabbrev]{cleveref}

\hypersetup{
    colorlinks=true,
    linkcolor=blue,
    filecolor=magenta,      
    urlcolor=blue,
    citecolor=blue,
    pdftitle={Overleaf Example},
    pdfpagemode=FullScreen,
    }

\theoremstyle{plain}

\theoremstyle{definition}

\theoremstyle{remark}

\usepackage[textsize=tiny]{todonotes}

\title{Optimizing Tensor Network Contraction Using Reinforcement Learning}
\author[1]{Eli A. Meirom}
\author[1]{Haggai Maron}
\author[1]{Shie Mannor}
\author[1]{Gal Chechik}
\affil[1]{NVIDIA Research, Israel}
\date{}
\begin{document}
\maketitle

\begin{abstract}
Quantum Computing (QC) stands to revolutionize computing, but is currently still limited. To develop and test quantum algorithms today, quantum circuits are often simulated on classical computers. Simulating a complex quantum circuit requires computing the contraction of a large network of tensors. The order (path) of contraction can have a drastic effect on the computing cost, but finding an efficient order is a challenging combinatorial optimization problem.

We propose a Reinforcement Learning (RL) approach combined with Graph Neural Networks (GNN) to address the contraction ordering problem.  The problem is extremely challenging due to the huge search space, the heavy-tailed reward distribution, and the challenging credit assignment. 
We show how a carefully implemented RL-agent that uses a GNN as the basic policy construct can address these challenges and obtain significant improvements over state-of-the-art techniques in three varieties of circuits, including the largest scale networks used in contemporary QC.

\end{abstract}


\section{Introduction}

Quantum computing has the potential to revolutionize the computer industry  \cite{gill2022quantum}, but quantum computers are still notoriously difficult to build \cite{franklin2004challenges,ladd2010quantum,cho2020biggest}. To devise and test quantum algorithms, researchers use classical computers to simulate the corresponding quantum circuits that implement them
\cite{huang2021efficient,gray2018quimb,itensor,patti2021tensorly}. In most cases,  simulating a complex quantum circuit involves computing a contraction of a huge network of tensors. 

Importantly, computing the contraction can be executed in a variety of ways, all yielding the same final result, but possibly at a very different computational cost. Section \ref{sec:formulation} explains this in more detail, but for now, a good analogy is the problem of computing the product of a sequence of matrices $A B C$. Both $(AB)C$  and $A(BC)$ yield the same result, but the computational cost may be very different depending on the matrix dimensions. 

Similarly, the cost of computing the contraction depends heavily on the order by which the tensors are contracted. Formally, the combinatorial optimization problem \textit{Tensor Network Contraction Ordering} (TNCO) attempts to determine the optimal order of contracting a network of tensors by iteratively contracting pair by pair \cite{pfeifer2014faster,xu2019towards,schindler2020algorithms}. 
TNCO can be naturally formulated using graphs, where nodes represent tensors and edges correspond to shared indices. Each edge is assigned a cost, which represents the computational cost of calculating the contraction along that edge. The objective is to find an ordering\footnote{we use path, sequence and ordering interchangeably throughout the paper.} over the pairwise contraction operations, or edges, that has the minimal total cost. 

Going beyond simulations of quantum computers, tensor network contractions play a key role in other areas of science, including many-body physics \cite{ran2020tensor}, statistical mechanics \cite{liu2021tropical}, and machine learning \cite{stoudenmire2016supervised}.

In the general case, TNCO is  computationally hard ~\cite{chi1997optimizing}, so designing efficient contraction algorithms remains a challenge. 
Some specific instances are computationally tractable, for instance, finding an  optimal order for matrix multiplication, when each tensor is a matrix \cite{cormen2009introduction}. For more complex networks, like those relevant for quantum circuits, there are multiple approaches \cite{pfeifer2014faster,xu2019towards,schindler2020algorithms,daniel2018opt}, employing a variety of tools from combinatorial optimization with different quality-complexity trade-offs. 
The current state-of-the-art 
is a graph-partitioning based solver \cite{gray2021hyper} that can produce high-quality solutions to networks with up to hundreds of tensors. For larger networks, the greedy approach (e.g. \cite{daniel2018opt}), which often results in poor contraction paths, remains the preferred choice.

\begin{figure}
    \centering
    \includegraphics[width=0.8\linewidth]{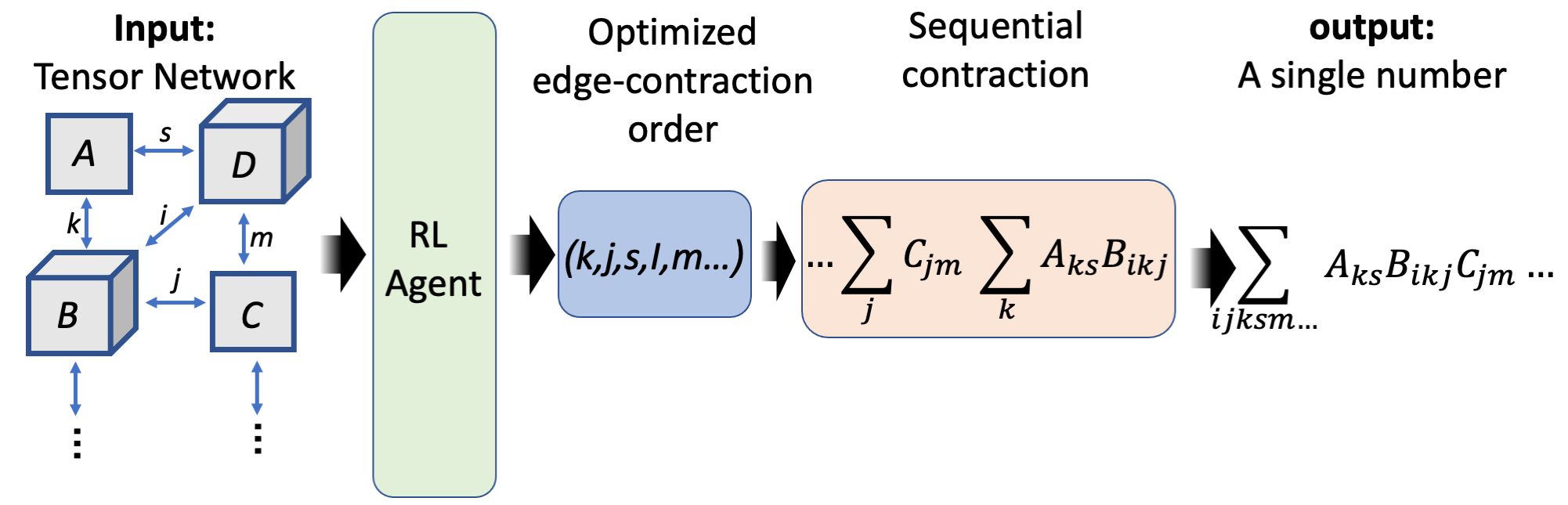}
    \caption{Contracting large tensor networks (sets of tensors with shared indices) is a major bottleneck when simulating  quantum circuits. The order by which summation takes place (blue) does not affect the output , but different orders may have  significantly different computation costs (orange). We tackle the Tensor Network Contraction Ordering problem: given a tensor network, our task is to find an optimal contraction order. To this end, we frame the problem as an RL problem and propose the first learning-based approach for solving it.} 
    \label{fig:fig1}
\end{figure}

Following the recent success of ML for combinatorial optimization \cite{dai2017learning,li2018combinatorial,sato2019approximation,almasan2019deep,nair2020solving,cappart2021combinatorial,peng2021graph,meirom2021controlling}, we devise a novel and effective Reinforcement Learning (RL) approach to solve TNCO. We formulate the problem as a Markov Decision Process (MDP): The state space is defined as a space of graphs, the action space is the set of edges to be contracted, and the transition function maps a graph to a contracted graph based on the selected edge. The reward function is defined as the (negative) cost of edge contraction. To solve this problem, an RL agent starts from a graph representing the input tensor network (TN) and is trained to sequentially select edges for contraction with the goal of minimizing the cumulative contraction cost . Importantly, at each step, the agent has to balance two, seemingly conflicting, objectives: (1) Select an edge with a low cost; (2) Select an edge whose contraction yields a graph that is efficiently contractable in the future.

Notably, applying an off-the-shelf RL approach to this formulation fails due to several unique challenges: (1) The search space is huge, relative to most RL applications; (2) The reward function is strongly heavy-tailed, breaking many assumptions of classical RL algorithms; and (3) The credit assignment problem becomes very challenging due to long episodes. Finally, it is not clear how non-learning based solvers can be incorporated into the training procedure. 

Aiming to address the challenges mentioned above, we introduce a novel RL approach, dubbed TNCO-RL. Our agent is implemented as a GNN \cite{battaglia2018relational} that takes as input the graph representation of the tensor network and outputs a distribution over its edges. This output distribution represents the relative likelihood that the agent selects a given edge.
The agent is trained to minimize the total contraction cost using a Proximal-Policy optimization (PPO) algorithm \cite{schulman2017proximal}. 

Our experiments indicate that TNCO-RL outperforms state-of-the-art baselines (e.g., graph-partitioning based methods) in a variety of experimental settings on both synthetic  tensor networks and tensor networks that originate from real quantum circuits \cite{arute2019quantum}. 
In summary, the contributions of this paper are: (1) We formulate the TNCO problem as a RL problem and discuss the main challenges that arise. (2) We propose a new approach to solve this problem using RL and GNNs. To the best of our knowledge, this is the first learning-based approach for the TNCO problem.
(3) We apply four technical components that allow to scale the RL approach: (a) path pruning, (b) optimistic buffer, (c) baseline integration, and (d) feature robustification, (4) we provide a training environment and benchmark to allow the community to further study this problem. 

\section{Previous work}
\paragraph{Tensor network contraction order optimization.} Finding efficient contraction paths for tensor networks was considered in several previous works. For tiny tensor networks consisting of up to 18 tensors, one can use brute force for finding the optimal path. \cite{pfeifer2014faster} showed that optimal solutions for slightly larger networks can be obtained when using a enhanced pruning techniques.  Greedy algorithms are a popular choice for larger problems (e.g., \cite{daniel2018opt}), often yielding sub-optimal paths. \cite{schindler2020algorithms} recently applied simulated annealing and genetic algorithms to the problem and showed that these methods outperform the greedy approach on small networks. \cite{gray2021hyper} survey several additional approaches and combine them with a hyperparmeter optimization approach. Moreover, they suggest using a graph partitioning algorithm \cite{schlag2016k, akhremtsev2017engineering} to guide the partitioning order. Their partitioning based solver is considered the current state of the art for Sycamore networks (several hundred nodes). \cite{modified_sycamore} propose a modified Sycamore
circuit TN, with a handcrafted specific ordering (rather than a general solver).

For certain specific network structures such as trees, optimal polynomial-time algorithms were suggested \cite{xu2019towards}.  \cite{huang2021efficient} improve the actual TN contraction under memory constraint, given a contraction order. They do not discuss how to initially find a good contraction order.

\paragraph{RL for combinatorial optimization.} Traditional algorithms often employ heuristics to solve combinatorial optimization problems. Machine learning, and RL in particular, can be used to \emph{learn} suitable rules from a specific data distribution at hand. Over the last few years, RL approaches have been applied to multiple combinatorial optimization problems such as the Traveling Salesman Problem (TSP) \cite{bello2016neural}, Max-Cut \cite{dai2017learning}, Bin Packing Problem (BPP) \cite{hu2017solving} , etc.  In particular, the combination of RL techniques with agents parameterized with Graph Neural Networks (GNNs) was shown to outperform the state of the art in several cases \cite{dai2017learning,almasan2019deep, meirom2021controlling}. We refer the interested reader to two recent surveys for more information \cite{yang2020survey, mazyavkina2021reinforcement}.   


\section{Background on tensor networks}\label{sec:formulation}

\begin{figure*}
    \centering
    \includegraphics[width=0.8\linewidth]{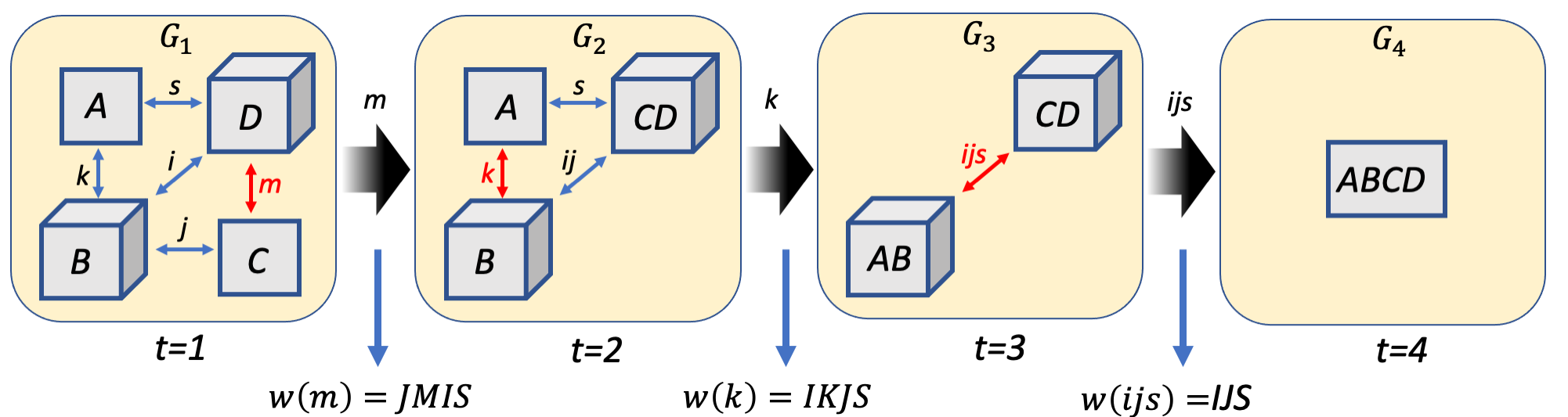}
    \caption{A sequence of pairwise contractions leading to a full contraction of the initial tensor network $G=G_1$. At each stage the red edge is contracted. $w$ represents the cost of each contraction step. }
    \label{fig:fig:sequential_contraction}
\end{figure*}

Here, we review the basic definitions of tensors and contractions. We refer the reader to \cite{bridgeman2017hand} for more information about TNs and their use in QC.

Given a set of tensors, tensor network contraction refers to the operation of summing over all shared indices. 
Tensor networks are usually sequentially contracted by contracting a pair of tensors at a time. In that case, the cost of computing the contraction depends heavily on the order by which tensors are contracted.  

\begin{wrapfigure}[10]{r}{0.26\textwidth}
\vspace{-25pt}
  \begin{center}
\includegraphics[width=0.18\textwidth]{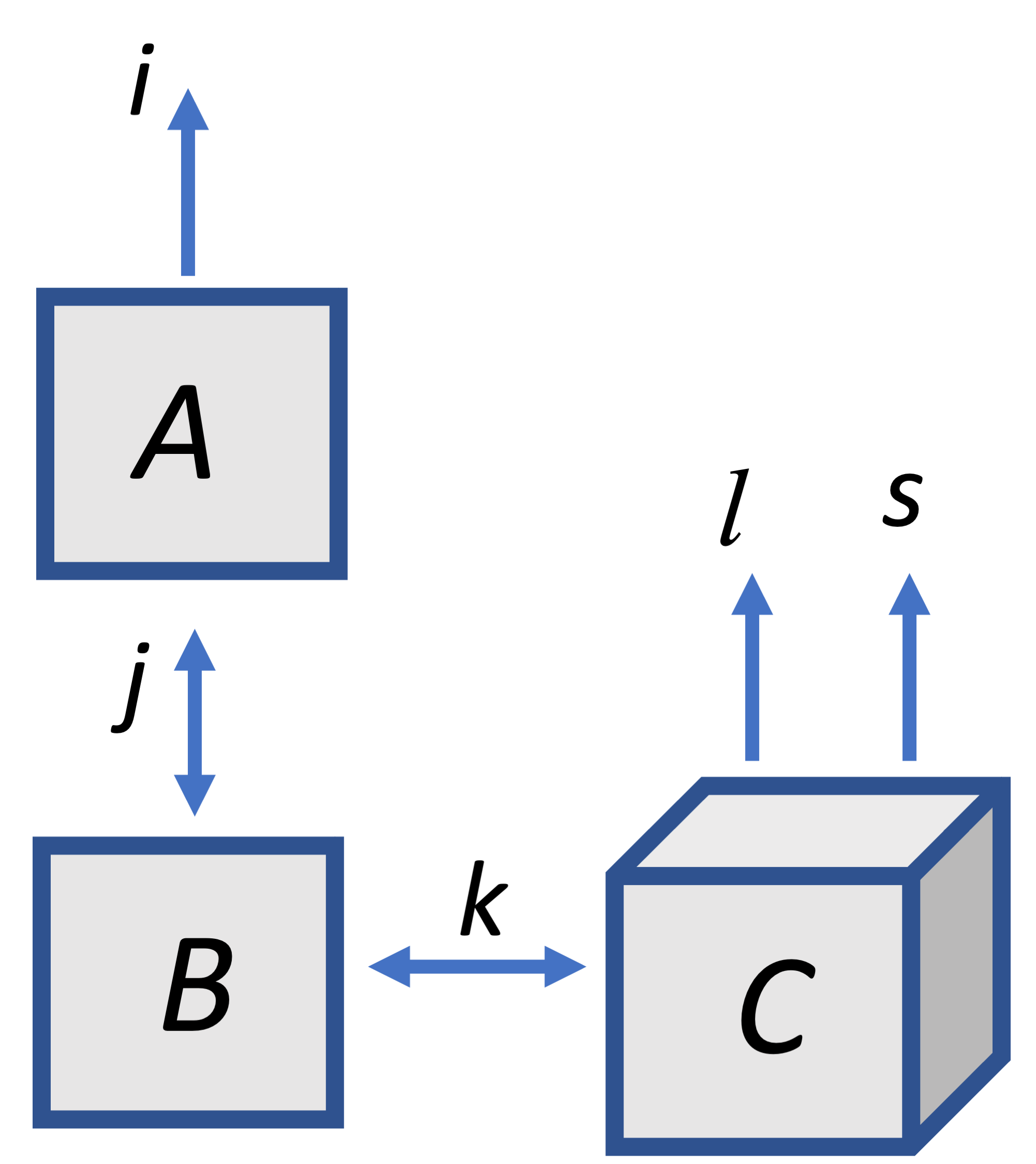}
  \end{center}
\end{wrapfigure}

Consider for example three tensors $A_{i,j}$,$B_{j,k}$, $C_{k,l,s}$ with corresponding dimensions $I, J, K, L$ (see inset). Their contraction is defined as a fourth tensor $T_{i,l,s}$ defined by: 
\begin{equation}\label{eq:contraction_example}
     T_{i,l,s}=\sum_{j,k}A_{i,j}B_{j,k}C_{k,l,s}
\end{equation}
If we first contract $j$ and then $k$, $IJK+IKLS$ flops are required to execute the contraction. However, if we first contract $k$ followed by $j$, the cost is $JKLS+IJLS$. For instance, if $I,L,S=2,J=10$ contracting the network with the first order costs $56$ flops, while contracting with the second order costs $160$ flops. With large tensor networks, the cost of different contraction orders (``paths''), can vary by orders of magnitude.

\paragraph{Tensors.} A tensor $A$ is a multidimensional array of numbers indexed by a subset of a finite symbolic index set $\mathcal{I}_A\subseteq\mathcal{I} = \{i,j,...  \}$. For each index $i \in\mathcal{I}$ we define a valid range, i.e. $i \in\{1,\dots,I \}$ called the \emph{extent} of the index. 

\emph{Pairwise Tensor contraction} is a common binary operation defined on pairs of tensors. Formally, the contraction of tensors $A,B$ is a new tensor $\mathcal{C}$, indexed by the symmetric difference of the index sets $\mathcal{I}_A \triangle \mathcal{I}_B:=(\mathcal{I}_A \cup \mathcal{I}_B) \setminus (\mathcal{I}_A \cap \mathcal{I}_B)$ whose values are defined by the following summation on the intersection of the index sets:
\begin{equation}\label{eq:pair_contr}
     \mathcal{C}_{\mathcal{I}_A \triangle \mathcal{I}_B} = \sum_{\mathcal{I}_A \cap \mathcal{I}_B} A_{\mathcal{I}_A}\cdot B_{\mathcal{I}_B}
\end{equation}



Here, with a slight abuse of notation, $A_{\mathcal{I}}$ denotes the indexing of a tensor by the indices in $\mathcal{I}$, and  $\sum_{\mathcal{I}}$ represents summation over all the indices in $\mathcal{I}$ (see the matrix product example below). In words, the tensor contraction operation is indexed by the indices that appear in only one of the tensors and its values are a summation over all shared indices. Matrix multiplication is a simple example of tensor contraction. In that case, assuming that a matrix $A$ is indexed by $\mathcal{I}_A =\{i,k\}$ and a matrix $B$ is indexed by $\mathcal{I}_B =\{k,j\}$ then the contraction $\mathcal{C}(A,B)$ is a tensor indexed by $\mathcal{I}_A \triangle \mathcal{I}_B =\{i,j\}$ and defined as:
$$\mathcal{C}(A,B)_{i,j}=\sum_{k=1}^K A_{i,k}B_{k,j},$$
which coincides with matrix multiplication.

\paragraph{Tensor networks.} A tensor network is a representation of a tensor contraction problem.  Formally, a tensor network is a collection of $n$ tensors $\{A^k\}_{k=1}^n$ where each tensor is indexed by subsets of an index set $\mathcal{I}_k \subseteq\mathcal{I}$. 
 
It is convenient to define a graph structure $G=(V,E,w)$ on a tensor network. This is done by first defining the set of nodes $V$ to be the set of tensors $\{A^k\}_{k=1}^n$. We then define the connectivity of the graph according to the shared index structure, namely $e_{kk'}\in E$ if $\mathcal{I}_k \cap \mathcal{I}_{k'}\neq \emptyset$. Lastly, we define a weight function on the edges $w:E\rightarrow \mathbb{R}_+ $. The weight function represents some measure of the time or space complexity of contracting the indices associated with the edge, and is usually a function of the relevant extents. In this paper, $w$ represents the number of flops needed to execute a contraction.\footnote{ The number of flops can be  calculated easily as it is proportional to product of the  extents of the indices that appear in both tensors.}

\paragraph{Tensor network contraction.} The contraction of a tensor network is a direct generalization of the binary contraction defined in Equation \ref{eq:pair_contr} above to a collection of tensors, namely:
\begin{equation}\label{eq:gen_contr}
     \mathcal{C}(A^1,\dots, A^n)_{\mathcal{I}_{\text{distinct}}} = \sum_{\mathcal{I}_{\text{shared}}} A^1_{\mathcal{I}_1}\cdot \dots \cdot A^n_{\mathcal{I}_n} 
\end{equation}
Here, $\mathcal{I}_{\text{shared}}$ is the subset of $\mathcal{I}$ which holds indices that appear twice in the collection $A^1,\dots, A^n$, and $\mathcal{I}_{\text{distinct}}=\bigcup \mathcal{I}_i\setminus \mathcal{I}_{\text{shared}}$  holds indices that appear once.\footnote{We assume that the total number of on index appearances is bounded by $2$. }

Executing the contraction in Equation \ref{eq:gen_contr} directly by using nested loops over all indices 
is severely inefficient due to the exponential complexity of that procedure. A more efficient option is to execute the contraction as a sequence of edge (or pairwise) contractions, $P=(e_1,\dots,e_{n-1})$. \footnote{this approach may use more memory for storing intermediate results, but often has better time complexity. }
Here, each contraction yields a new graph $G_t$, starting with $G_1=G$. At each step $t$, a pair of tensors connected by the edge $e_t=(u,v)$ is contracted according to Equation \eqref{eq:pair_contr}. Then a new tensor network $G_{t+1}=(V_{t+1},E_{t+1},w_{t+1})$ is defined by removing the two contracted tensors $u,v$ and adding the resulting tensor $T_{t+1}$  to $V_{t+1}$. The adjacency structure is defined according to shared indices as described above (i.e.,  all edges that have previously been connected to either $u$ or $v$ will now be connected to $T_{t+1}$ instead), see Figure \ref{fig:fig:sequential_contraction}.


\section{TNCO: A definition and a RL formulation}

We now define the TNCO problem and formulate it as an RL problem. We first motivate our approach, then formulate TNCO as an MDP, and discuss two possible learning setups that are of interest.

\subsection{Motivation}

Recently, reinforcement learning has been shown to be a promising approach for solving combinatorial optimization problems \cite{yang2020survey, mazyavkina2021reinforcement}. The main idea is to break down the assignment of variables in the problem to a sequential process and formulate it as a Markov Decision Process. If the problem instance at hand is similar to other instances, a learning approach may succeed in generalizing past experience and problems and achieving a good solution.

Many combinatorial optimization problems can be formulated using a graph structure. In particular, TNCO can always be written as a graph under the common assumptions in the literature \cite{gray2021hyper}. The most successful current approaches for TNCO use graph partitioning tools \cite{gray2021hyper}. Graph Neural Networks (GNNs) have been shown to learn useful representations of graphs \cite{ZHOU202057}. Therefore, it makes sense to combine their representative power with an RL framework.

As an important alternative, Genetic Algorithms (GAs)  have been considered for the TNCO problem (e.g. \cite{schindler2020algorithms}). Unfortunately, they suffer from two main drawbacks. First, the search space is huge, requiring very large generations, and that makes GAs very slow and limits them to TNs of a few tens of nodes. Second, GAs do not encode the structure of the graph, so the similarity between graphs from the same distribution is not used.

\subsection{The TNCO problem} 
Given a tensor network represented as a graph $G=(V,E,w)$, a contraction path $P=(e_1,\dots,e_{n-1}), \ e_t \in E_t$ and a corresponding sequence of graphs $(G_1,\dots G_n)$, we define the total contraction cost $c$ as the sum of weights of the contracted sequence: 
\begin{equation}
    c(P)=\sum_{t=1}^{n-1} w_t(e_t)
\end{equation}


The goal in TNCO is to find a sequence of edge contractions $P$ of minimal cost, namely:

\begin{subequations} \label{eq:objective}
\begin{align}
        & P^*(G) = \text{argmin}_P ~~ c(P) \\
        & \mathrm{s.t.}  \quad   P=(e_1,\dots, e_{n-1}), e_t\in E_t
\end{align}
\end{subequations}

\subsection{MDP Formulation}


To formulate TNCO as an MDP, we specify the state and action spaces, as well as the transition dynamics and the cost function. We use the cost rather than the reward to adhere to the QC community standards. To define an MDP, we need to define the state space $\mathcal{S}$; the action space $\mathcal{A}$; the transition function $\mathcal{F}$; and the immediate reward/cost function $\mathcal{R}$. We discuss the design choice of the various elements below. 

The state space $\mathcal{S}$ is the space of all weighted graphs,  $\mathcal{S}= \{ G=(V,E,w) \}$. This is a very large space, but it provides a sufficient statistic to evaluate the dynamics of every action. Also, by choosing such a state space, we open the possibility for generalization over states, that is, generalization over graphs.
The actions available in a given state is the set of edges $E$, which naturally varies from one graph (state) to another. 
The transitions in this problem are deterministic: $\mathcal{F}(G,e)$ takes as input the current state $G$ and an edge $e\in E$, and outputs a new graph that represents the tensor network after contracting the edge $e$ as explained in Section \ref{sec:formulation}. 
The immediate cost $\mathcal{R}(G,e)$ for an action $e$ is the weight of the edge $w(e)$. The agent's objective is to find the best contraction sequence, that is, to minimize the overall cost. 
\subsection{The learning setup: Single and multi- network learning}
We consider two learning setups.  First, training an agent on a single network. This setup is suitable for optimizing large networks, which is typically done once, as in quantum-supremacy tests. The agent would repeatedly contract the same network and learn to find better contraction paths. 
The second setup involves training the agent on a set of training networks and then applying it at inference time to new networks that are somewhat similar to the training networks, for example, taken from the same distribution. 

\begin{figure}
    \centering
    \includegraphics[width=0.5\linewidth]{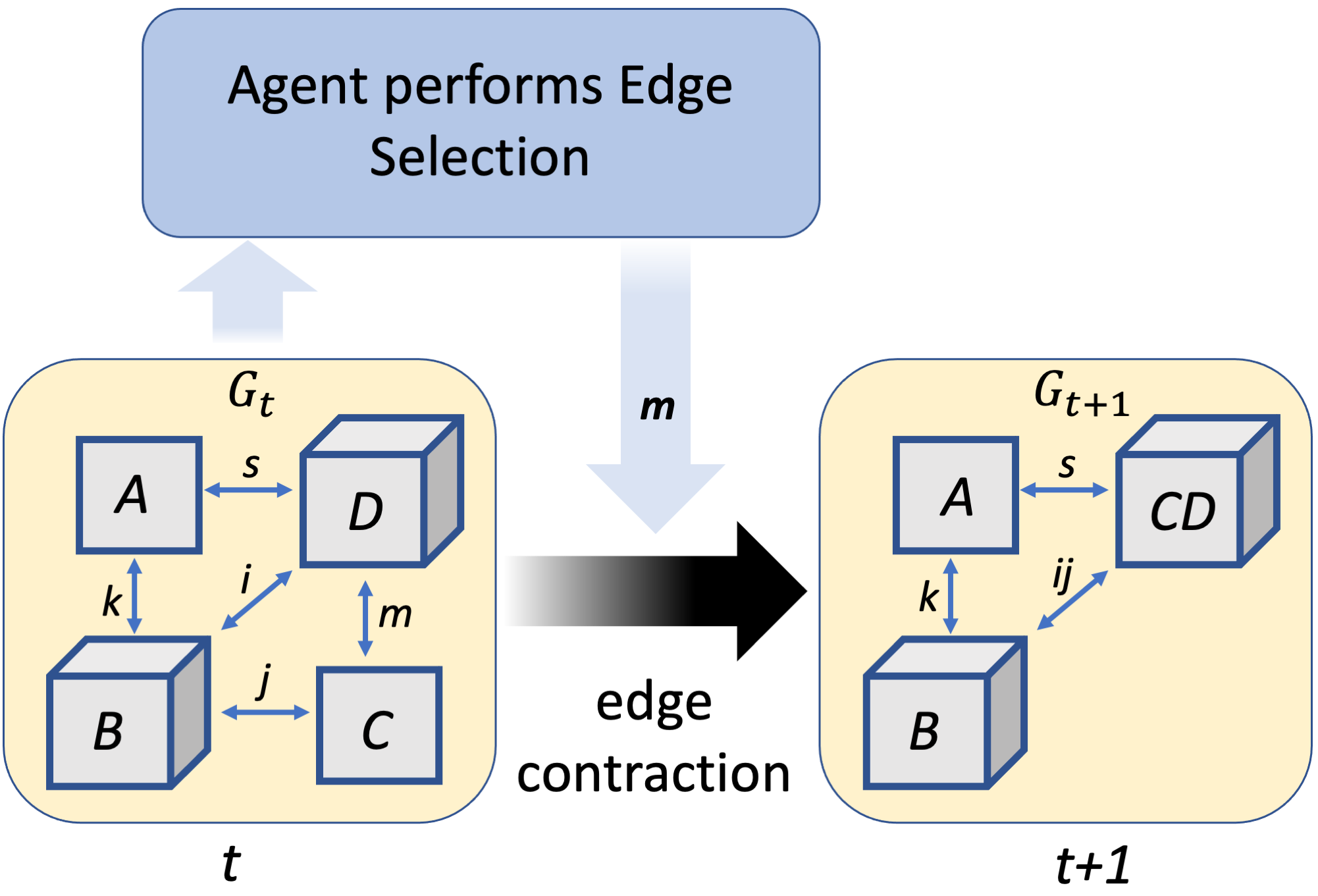}
    \caption{ Our goal is to find an agent $H$ that takes the current graph $G_t$ as input and selects the next edge to contract, such that the sequence of selected edges $P$ has optimal cost $c(P)$.}
    \label{fig:agent}
\end{figure}

\paragraph{Problem 1: Learning in a single network.} Given a tensor network represented as a graph $G$ we look for an agent $H$ in the class of agents $\mathcal{H}$ whose edge contraction choices yield an optimal contraction sequence. Formally, the agent should solve the following optimization problem:
\begin{subequations} \label{eq:objective}
\begin{align}
        & \text{argmin}_{H\in \mathcal{H}} ~~ c(P) \\
        & \mathrm{s.t.}  \quad  P=(e_1,\dots, e_{n-1})\\
        & e_{t}=H(G_t),\quad t=1,\dots,n-1
\end{align}
\end{subequations} 
Figure \ref{fig:agent} illustrates the edge selection process. 

\paragraph{Problem 2: Learning and generalization to new networks}
Here, an agent is trained on multiple input graphs $\{G^i\}_{i=1}^m$ sampled from an underlying tensor network distribution. The agent is tested on unseen target graphs from the same distribution. In this case, the training objective is as follows:
\begin{subequations} \label{eq:objective_learning}
\begin{align}
        & \text{argmin}_{H\in \mathcal{H}} ~~ \sum_{i=1}^m c(P^i) \\
        & \mathrm{s.t.}  \quad  P^i=(e^i_1,\dots, e^i_{n-1})\\
        & e^i_{t}=H(G^i_t),\quad t=1,\dots,n^i-1
\end{align}
\end{subequations} 
This setup is suitable if one is interested in quickly solving TNCO instances from the same distribution. Once a model has been trained, it can be applied very quickly to infer paths for new input tensor networks. 

\section{Challenges in Applying RL to TNCO \label{sec:chal}}
Applying RL to TNCO raises five challenges that need to be addressed. Section \ref{sec:appproach} describes our concrete solutions to these challenges in the context of TNCO.

\begin{wrapfigure}[18]{r}{0.45\textwidth}

\vspace{-50pt}
  \centering
\includegraphics[width=0.42\textwidth]{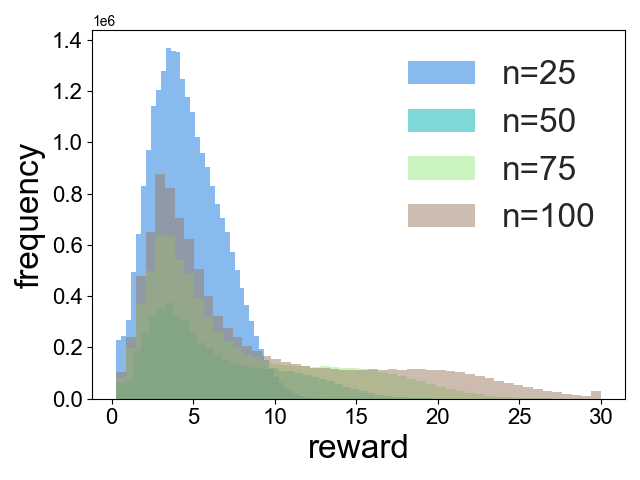}
\vspace{-10pt}
\caption{\footnotesize Reward distribution of random actions in random TNs. \label{fig:reward_distributation}} 
\end{wrapfigure}

\paragraph{Challenge 1: Wide dynamic range.} 
The contraction costs and tensor sizes typically vary across many orders of magnitude.

Deep neural networks are hard to train when costs are highly variable. Importantly, naive normalization does not address the problem. When the distribution of costs remains strongly unbalanced, normalizing by any constant quantity, like the max or median, washes out the signal in (at least) one end of the feature-value distribution. The inset Fig. \ref{fig:reward_distributation} shows a histogram of edge contraction costs (in log10 scale) sampled from random paths on random TNs with $n\in \{25,50,75,100\}$ tensors. 


\paragraph{Challenge 2: Huge search space.} 
The number of possible contraction paths in our problem is enormous. 
Even among very large action and state spaces, the TNCO problem is an outlier. For comparison, the average length of an episode in a game of GO \cite{silver2017mastering} is 200, and so is the action space (branching factor). In a TNCO problem based on a modern quantum circuit, the action space and the length of an episode can reach $O(1000)$ steps and actions. Therefore, the number of possible trajectories in TNCO is \emph{hundreds of orders of magnitude larger} than that of GO. 


\paragraph{Challenge 3: Slow convergence due to Heavy-tailed cost distribution.}
Nearly all research in RL is based on the assumption that the immediate cost has a sub-Gaussian distribution. For approaches based on stochastic approximation, like Q-learning or policy gradient, a bounded second moment is needed. 
In TNCO, as shown above for Challenge 1, the reward seems to have a heavy-tailed behavior, with its empirical distribution resembling a log-normal distribution. This implies that the variance is expected to be very large even if finite. Moreover, this phenomenon becomes worse as the networks become larger.
We note that previous results for multiarmed bandits and RL \cite{bubeck2013bandits,zhuang2021no} have shown that if the reward distribution is not sub-Gaussian, the convergence time is $|\mathcal{S}|^\alpha, \alpha>1$, where $\alpha$ is a function of the highest moment of the distribution that converges. Consequently, the TNCO problem suffers from a huge state space and a heavy-tailed reward distribution, and therefore vanilla policy gradient algorithms are unlikely to converge in a reasonable time.


\paragraph{Challenge 4: Incorporating existing solvers.} 
Existing meta-heuristic approaches and learning-based approaches have complementary strengths. Existing solvers can often find the optimal solution for  small problems. Learning-based methods usually only approximate the optimal solution, but work well on large problems. The challenge is to enjoy the benefits of the two approaches.

\paragraph{Challenge 5: Credit assignment problem.} 
In TNCO, as in other combinatorial optimization problems, early decisions can have a crucial effect on the overall trajectory cost. TNCO's lengthy episodes compared to many other RL problems exacerbate the credit assignment problem.

\section{Our Approach: RL-TNCO} \label{sec:appproach}

We start with a short overview, and then discuss the graph representation and processing module in our approach. Finally, we review the RL framework and focus on how we address the challenges discussed in the previous section.  

\paragraph{Overview.} As the basis for our RL framework, we use PPO \cite{schulman2017proximal}, though the choice of the specific policy search algorithm is less important here.
Tensor networks are represented as graphs, and we use a GNN to parameterize our policy \cite{battaglia2018relational}. At each step, a graph representation of the current tensor network $G$ is fed into the GNN, and the GNN outputs a distribution over the edges to be selected. An edge is then sampled from the distribution and contracted.  This yields a smaller tensor network with one less node, represented by the graph $G'$. The weight of the selected edge is added to the total path cost, which is used to calculate the loss. The process is repeated until no edges remain in $G'$.  At inference time, we treat the agent output as a distribution over contraction sequences. We sample $\ell$ sequences, edge by edge, and use the sequence that yielded the minimal total contraction cost. 

\subsection{Graph representation and architecture } We represent the graph as a tuple $(A,X,E,g)$ where $A\in \{0,1\}^{n\times n}$ is the adjacency matrix of the tensor network, $X\in \mathbb{R}^{n\times d}$ is a node feature matrix that holds features for each tensor, $E\in \mathbb{R}^{m\times d'}$ is an edge feature matrix, and $g\in \mathbb{R}^{d''}$ is a global graph feature vector. Here, $m$ is the number of edges in the tensor network and $d,d',d''$ are arbitrary feature dimensions. We experimented with several choices of initial features and converged on the following choice: For node and global features we use constant input features, i.e. $X=[1,\dots,1],~g=[1]$. The edge  features of an edge $e=(u,v)$ are: (i) its are the contraction cost; (2) the edge's extension (product of the extents of the associated indices), and (3) the  greedy score of the edge  $e=(u,v)$, based on a heuristic approach in OPT-Einsum \cite{daniel2018opt} $\prod_{i\in\mathcal{I}_u}I(i) + \prod_{i\in\mathcal{I}_v}I(i) - \prod_{i\in\mathcal{I}_u \triangle {I}_v}I(i)$. Here, $I(i)$ is the extent of index $i$.

As we have both node and edge features, we choose to use a popular message passing GNN model  \cite{battaglia2018relational} (Algorithm 1, pp. 12) that maintains learnable node, edge and global graph representations. The model is composed of several message passing layers, $L^k\circ\dots \circ L^1$ where each $L^i$ updates all the representations, i.e.:
$$ X^{i+1},E^{i+1},g^{i+1} = L^i(A,X^i,E^i,g^i;\theta^i),$$
Each layer $L_i$ updates the features sequentially: node and edge features are updated by aggregating local information, while the global feature is updated by aggregating over the whole graph. We denote the parameters of the MLPs that are used in a layer $L_i$ as $\theta_i$, and note that these are the only learnable parameters in the model. At the last layer $i=k$ we use a single dimension for edge features, that is, $d'=1$ and apply a normalization function over edge scores $s(e)$ \cite{meirom2021controlling},  $$\Pr(e) = \frac {s'(e)}{\sum_e s'(e)},\quad s'(e)=s(e)-\min_{e'} s(e')+\epsilon.$$ This provides us with distribution $\Pr(\cdot)$ over the edges, which is used for sampling an edge. The critic module shares the same architecture, on top of which a final pooling layer over the nodes is added to obtain the value function.

\subsection{Algorithmic modifications}
To address the challenges discussed in Section \ref{sec:chal}, we incorporated several algorithmic modifications to vanilla deep policy gradient algorithms. 


\paragraph{Path pruning.}
To restrict the huge search space (Challenge \#2), we use path pruning. All the costs in a TNCO problem are strictly positive. Therefore, if the cost accumulated at some point in a path exceeds the smallest observed cost $C^{min}$, this path cannot be optimal and can be pruned. 

Unlike inference, using path pruning in training cannot be done in a naive way.  
First, if we prune paths during the path-collection phase, the vast majority of paths get pruned and too little data is left for training. A better approach would be to use the path up to the termination point. In this approach, we need to estimate the cost of that path suffix, and this requires estimating the future expected return at the termination state, which is commonly done using the value function. However, estimating that value itself relies on the assumption that we can collect enough full (completed) trajectories, so the critic can learn to approximate the value at the tail of the trajectory. Since in TNCO most trajectories terminate prematurely, the approximation of the value function at the termination state becomes very poor. To mitigate these issues, we instead approximate the termination value after $l$ contractions by 
$\min C(G,e) + ( \frac{C_l}{l} + \frac{C^{\min}}{n} )\cdot \frac{n-l}{2}$. The last term averages the mean known cost per step 
and the mean cost per step of the current trajectory $\frac{C_l}{l}$.

\paragraph{Optimistic buffer.}
In our setup, good paths are very rare. Even when the agent encounters a good path, it is often overwhelmed by poor trajectories, and the information is often washed out by negative path updates, leading to slow convergence (Challenge \#3). Furthermore, deep policy gradient algorithms often apply gradient clipping, limiting the effect of the highly valuable but very rare trajectories. To overcome this and address the slow convergence rate of the PG algorithm in our setup, we add an off-policy element to the on-policy PPO algorithm -- an optimistic buffer.

This buffer contains tuples $(G,w(e),e,C(G,e))$\footnote{In standard RL notation, $G=s, w(e)=r, e=a.$}, where $C(G,e)$ is the total empirical accumulated cost from this state after performing the action $e$, as logged from a specific trajectory.  This is a pessimistic estimation of the optimal value function $V^{opt}(G)$, as the optimal value function is the maximal accumulated reward over all trajectories from $G$. 
After each training step, we update this buffer. For each sample such that $V(G)>C(G,e)$ from either the roll-out buffer or the optimistic buffer we assign a score $V(G)-C(G,e)$, while for all other samples we assign a score of $0$, effectively eliminating them. We then sample $b$ samples according to the normalized score distribution, where $b$ is the optimistic buffer size.
Then, we update the actor and critic modules using PPO loss with samples drawn from this buffer. This results in an off-policy bias, which shifts both the actor (and critic) towards the optimal policy.

\paragraph{Robust feature scaling.}
To address the problem of a wide dynamic range from Challenge \#1, we propose a novel scheme of feature robustification.  We apply two transformations to each input feature that assign high dynamic range to the tails. In addition, we apply a logarithmic transformation that compresses the dynamic range of the heavy-tail distribution.

These transformations are applied to every feature. For a specific feature $z$, denote the vector $z$ values across all nodes as $y_z$. This is a column vector of the edge feature matrix $E$. If all the values in $y_z$ have the same sign, the scale feature is $ \log(abs (y_z) )$, otherwise it is $\log(y_z-\min(y_z)+1)$.  

Next, we apply the two transformation that enhance the dynamic range at the distributation tails.  We use the median $m$ of $y_z$ as a ``border" to separate high and low ranges. We scale each column as $\frac{y_z - m}{h(y_z)-m}$, where $h(\cdot)$ is the scale of the high (low) ranges, e.g. $h(\cdot)= \max(\cdot),\min(\cdot)$. Then, we clip the high and low regimes to $[-100,100]$.


\paragraph{Leveraging existing solvers.}
To benefit from existing algorithms (Challenge \#4), which can often find optimal solutions in small networks,
we tested 
two approaches to incorporate existing solvers. (1) For a solver based on action scores (such as a greedy approach), we incorporated its score as additional features that the algorithm can learn to use. 
(2) Use a fast solver to optimize part of a problem. For example, when solving a large TNCO problem, use a greedy approach to find a prefix of the contraction path, using the low-cost contractions, and then use RL to find the rest. In the later approach, RL spends optimization cycles on the most important (heavy) late stage contractions . 

\begin{wrapfigure}[10]{r}{0.4\textwidth}
\vspace{-25pt}
  \begin{center}
\includegraphics[scale=0.38]{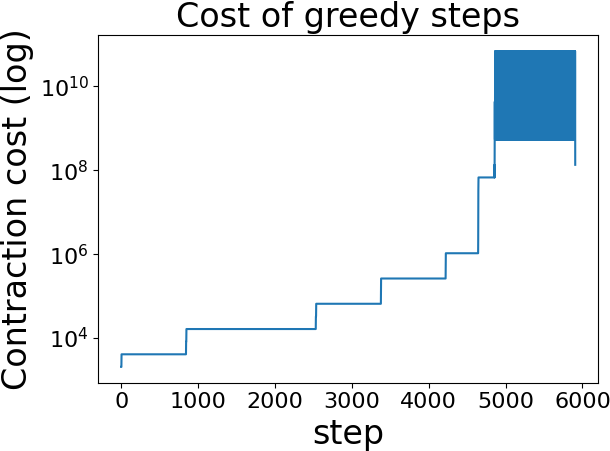}
    \caption{\footnotesize The contraction cost per step of the greedy algorithm on the max-cut TN.}
    \label{fig:greedy_cost_per_step}  \end{center}
\end{wrapfigure}

The intuition behind using greedy at the prefix is that the mean cost of contraction grows during optimization, hence most headroom to improve is at the suffix. For example, Fig. \ref{fig:greedy_cost_per_step} presents the cost per step of the greedy solver for the max-cut TN. 90\% of the cost incurred during the last 100 steps, while 95.5\% of the contraction cost is incurred during the last 1000 steps.

We also implemented applying the solver at the suffix, but since the prefix graph is very large (10K edges), only small batches fit in memory even on a 40GB GPU. As a result, applying RLTNCO to the prefix is, unfortunately, slower by orders of magnitude. Furthermore, 

The credit assignment problem (Challenge 5) is addressed through the use of a general-purpose RL scheme and a lot of compute power. Developing an algorithmic approach to address this challenge is an open problem. 

\section{Experiments}
\subsection{Setup} 
We conducted experiments on three types of networks: synthetic networks ($\leq 100$ tensors), Sycamore circuit networks ($100-400$) and Max-cut ($>5000$) networks. The latter two networks originate from real quantum circuits. 

When multiple tensor networks drawn from a shared distribution are available, it is beneficial to train a model to solve circuits from that distribution. Synthetic networks allowed us to test how well our method generalizes to new networks from the same distribution. Therefore, we applied the multi-network setup on the synthetic networks.  

The Sycamore and Max-cut are currently on the frontline of the quantum supremacy regime, and are often simulated in supercomputers. As such, the key challenge here is to reduce the total flop count, whereas the runtime of the TNCO solver is of less concern. The number of both Sycamore circuits and max-cut graphs is very low (four and one, respectively), and therefore we used the single-network approach in these tensor networks.

\paragraph{Baselines.} We compare our approach to three baselines: \textbf{(1) OE-GREEDY:} the greedy solver from OPT-Einsum \cite{daniel2018opt}. This solver provides a very fast solution which is usually sub-optimal. It runs for a few milliseconds for small networks, a few seconds for huge ones; \textbf{(2) CTG-Greedy:} the greedy solver from \cite{gray2021hyper}; and \textbf{(3) CTG-Kahypar,} a strong graph partitioning  based solver from \cite{gray2021hyper}. Both  CTG-Greedy and  CTG-Kahypar run with a time limit as specified in each experiment. Since OE-Greedy is very efficient, if another algorithm has a worse solution, we replace it with the OE-Greedy solution.

\paragraph{Implementation and system.} We used NVIDIA DGX-V100 for all experiments. We used the pyTorch Geometric \cite{fey2019fast} for GNN implementations and Stable Baselines \cite{raffin2019stable} for the RL framework.

\paragraph{Hyper-parameter search.} We conducted a hyper-parameter search on the following choices: (1) Number of GNN layers $k\in \{3,4,6\}$ ;(2) Using PairNorm or not  \cite{zhao2019pairnorm} ;(3) Neighborhood aggregation type (max/mean/sum) ;(4) Edge normalization method (softmax vs. the approach from \cite{meirom2021controlling}).

\begin{figure}[t]
    \centering
    \includegraphics[width=.45\linewidth]{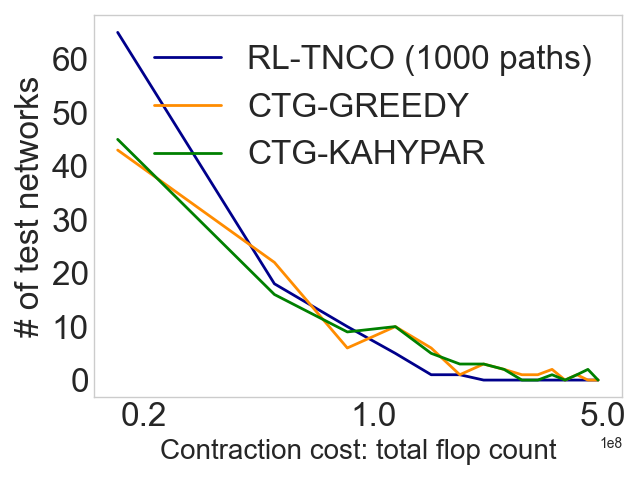}
    \includegraphics[width=.45\linewidth]{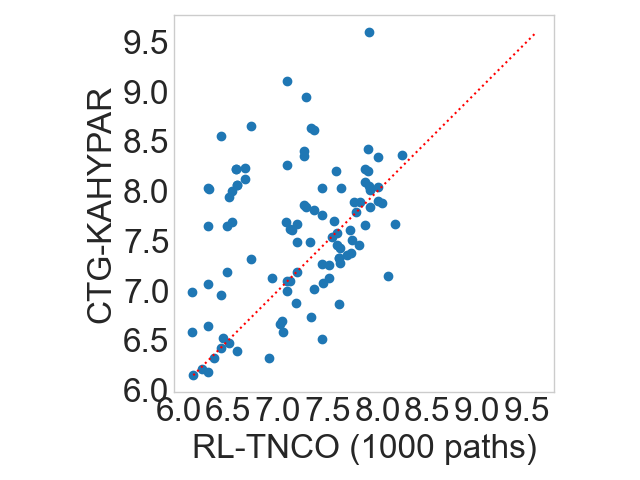}
    \caption{Contraction costs for synthetic networks of $n=50$, scale is $10^8$. (left) A histogram of the total flop count. (right) Each point corresponds to a network from the test set, axes correspond to contraction costs for RL-TNCO and Khaypar.  
    }
\label{fig:small_scale}
\end{figure}

\setlength{\tabcolsep}{3pt}
\begin{table*}[ht]
  \centering
  \small
  \caption{Total flop count in synthetic networks of various sizes.}
    \begin{tabular}{l|llll|llll}
    \toprule
          & \multicolumn{4}{c}{Mean $\pm$ std} & \multicolumn{4}{c}{Median $\pm$ MAD} \\
    \midrule
    $n$ & \multicolumn{1}{l}{25} & \multicolumn{1}{l}{50} & \multicolumn{1}{l}{75} & \multicolumn{1}{l}{100} & \multicolumn{1}{l}{25} & \multicolumn{1}{l}{50} & \multicolumn{1}{l}{75} & \multicolumn{1}{l}{100} \\
    \midrule
    Scale & $10^4$ & $10^7$ & $10^{10}$ & $10^{12}$ & $10^4$ & $10^7$ & $10^{10}$ & $10^{12}$ \\
        \midrule
    OE greedy & 53.7 $\pm$ 83.7 & 75.4 $\pm$ 253 & 104 $\pm$ 410 & 5296 $\pm$ 38310 & 27.7 $\pm$ 15.6 & 11.3 $\pm$ 10.0 & 4.5 $\pm$ 4.4 & 26.4 $\pm$ 25.4 \\
    CTG greedy & 40.3 $\pm$ 70.9 & 12.8 $\pm$ 45.6 & 8.3 $\pm$ 27.1 & 27.9 $\pm$ 103.0 & 20.3 $\pm$ 9.6 & 4.2 $\pm$ 3.4 & 0.9 $\pm$ 0.8 & 2.2 $\pm$ 2.0 \\
    CTG kahypar & 46.4 $\pm$ 76.5 & 13.4 $\pm$ 42.2 & 4.1 $\pm$ 18.2 & 54.2 $\pm$ 463 & 24.8 $\pm$ 12.4 & 4.3 $\pm$ 3.5 & 0.4 $\pm$ 0.4 & \textbf{1.2 $\pm$ 1.1} \\
    \midrule
     RL-TNCO (50) & 13.9 $\pm$ 7.3 & 3.9 $\pm$ 4.5 & 1.6 $\pm$ 4.6 & 9.1 $\pm$ 21.1 & 13.5 $\pm$ 7.3 & 2.0 $\pm$ 4.5 & 0.2$\pm$ 4.6 & 2.3 $\pm$ 21.1 \\
    RL-TNCO (1000) & \textbf{13.1 $\pm$ 6.5} & \textbf{3.2 $\pm$ 3.5} & \textbf{1.2 $\pm$ 2.7} & \textbf{5.5 $\pm$ 11.0} & \textbf{12.5 $\pm$ 3.2} & \textbf{1.8 $\pm$ 1.5} & \textbf{0.2 $\pm$ 0.1} & 1.8 $\pm$ 1.6 \\
    \bottomrule
    \end{tabular}%
  \label{tab:small_scale}%
\end{table*}%

\begin{table}[ht]
  \centering
  \caption{Inference time comparison in seconds for synthetic networks.}
     \begin{tabular}{lrrrr}
     \toprule
     \# nodes      & 25    & 50    & 75    & 100 \\
     \midrule
     OE greedy & 0.02 s& 0.07 s& 0.1  s & 0.06 s\\
     CTG greedy & 300 s  & 300 s  & 300 s  & 300 s\\
     CTG kahypar & 300 s  & 300 s  & 300 s   & 300 s\\
     \midrule
     RL-TNCO (50) & 1.6 s  & 4.2 s   & 6.8 s  & 9.6 s\\
     RL-TNCO (1000) & 33 s & 74.6 s & 124.2 s & 162.4 s\\
     \bottomrule
     \end{tabular}
  \label{tab:timing_small}
\end{table}

\subsection{Results}
\paragraph{Synthetic networks.} We first applied our approach to synthetic networks with sizes $n=25$, $50$, $75$ and $100$ nodes. Following \cite{gray2021hyper} we generate graphs using the OPT-Einsum  package \cite{daniel2018opt} with an average degree $d=3$ and tensor extents that are sampled i.i.d.~from a uniform distribution on $\{2,3,4,5,6 \}$. In all experiments, we train on randomly generated tensor networks and test on a set consisting of specific $100$ equations. For each equation, we sampled numerous paths, and report the best found path. We varied (and report) the number of paths sampled at test time $\ell \in \{ 50,1000\}$.  This allows us to trade off runtime with solution quality. 

We repeat each experiment with three random restarts (seeds) and report the average solution. CTG-Kahypar and CTG-Greedy ran for 300 seconds for each TN. 

The results are reported in Table \ref{tab:small_scale}. RL-TNCO outperforms all baselines by a large margin ($2\times$--$5\times$), even when sampling only $\ell=50$ sampled contraction sequences. Figure \ref{fig:small_scale} (top) provides a more detailed analysis of the results showing that the RL-TNCO algorithm tends to provide more efficient solutions. Figure \ref{fig:small_scale} (bottom) 
shows a paired comparison of the results of RL-TNCO and CTG-Kahypar. 
RL-TNCO is better than CTG-Kahypar in 72\% of the networks (points below the line). 
Table \ref{tab:timing_small} compares the running time of all algorithms at inference time. RL-TNCO runs significantly faster than the CTG baselines. 

Finally, we tested how RL-TNCO models trained on tensor networks of one size perform when tested with tensor networks of a different size. This \textit{size-generalization} problem is an active research field \cite{yehudai2021local}
%
We trained four models on networks with different fixed sizes, having 25, 50, 75, and 100 nodes. We tested how each model performs on a test set containing TNs with 25 nodes. We find that models trained on TNs with 50, 75, and 100 nodes reached $99.7\%,94.1\%,91.8\%$ of the performance of ``same-size" models, trained on TNs of size 25. This shows that performance deteriorates slightly when the train distribution deviates from the test distribution.


\paragraph{Sycamore networks.} We tested our framework on four circuits based on the sycamore processor \cite{arute2019quantum}, with 53 qubits and $m=10$,$12$,$14$,$20$ cycles. These networks had 162-379 tensors after simplification using the QUIMB library \cite{gray2018quimb}. \footnote{With this simplifications, the circuits are not identical to those reported in \cite{gray2021hyper}.} We consider the single-circuit learning setup. We ran CTG-GreedyGreedy and CTG-kahypar for 3 hours on each circuit. RL-TNCO outperforms the baselines in three out of four TNs.
See Table~\ref{tab:medium_scale} for a comparison of our results to the baselines, and Table \ref{tab:m20_analysis} for a quantification of gains for Sycamore $m=20$

\begin{table}[t]
  \centering
     \small
  \caption{Contraction cost for Sycamore networks of four sizes, with $m=10, 12, 14, 20$. Values indicate flops count.}
    \begin{tabular}{lrrrr}
    \toprule
    & \multicolumn{1}{l}{m=10} & \multicolumn{1}{l}{m=12} & \multicolumn{1}{l}{m=14} & \multicolumn{1}{l}{ m=20} \\
    \midrule
    \# Tensors  &    162   &    210   &   244    & 379 \\
    \midrule
    Scale & $10^{10}$ &  $10^{12}$ & $10^{14}$ &  $10^{18}$  \\
    \midrule
    OE greedy &  $5.71\cdot 10^{4}$     &  $2.96\cdot 10^{8}$     &  $1.52\cdot 10^{4}$     &  $2.04\cdot 10^{13}$ \\
        CTG greedy &  3.78     &  102     &  19.2     & 8.57  \\
    CTG kahypar &  \textbf{2.01}    &  43.6    &  5.07    &  5.83 \\
    \midrule
    RL-TNCO  &  5.44     &   \textbf{7.4}     &  \textbf{2.63}     & \textbf{3.50} \\
    \bottomrule
    \end{tabular}%
  \label{tab:medium_scale}%
\end{table}%

\begin{table}[t]
  \centering
     \small
  \caption{Quantification of time, money and energy saved with RL-TNCO for Sycamore m=20. Assuming GPU utilization of 0.5, MWh price 100\$, Tera flops per second of 19.5 on an NVIDIA A100 card (fp32/fp64), and using $1M$ repeats. Reducing the number of shots is an active research field \cite{pan2021solving}.}.
    \begin{tabular}{lcc}
    \toprule
    & RL-TNCO & Best baseline \\
    \midrule
    Path optimization time (hours)    & 24   & 24   \\
    Contraction cost ($10^{18}$flops) & 3.49 & 5.83 \\ 
    Single contraction time (hours)   & 99.43 & 166.43 \\
    Total time (K hours)              & 99,430 & 166,096 \\
    \midrule
    Time saved (K hours)              & 66,666 &  \\
    Mw saved                          & 60,000,000 &  \\
    Money saved (M\$)                 & 6,000 &  \\
    \bottomrule
    \end{tabular}%
  \label{tab:m20_analysis}%
\end{table}%

\paragraph{Max-cut networks.} We applied RL-TNCO to a tensor network that originated from a Max-Cut problem \cite{patti2021variational}. The circuit consists of $1688$ qubits with depth 21, yielding a network with 5908 tensors. Due to the large network size and as discussed in Section \ref{sec:appproach}, we first execute a greedy solver and use RL-TNCO for optimizing the last $k$  contraction steps. Table \ref{tab:large_scale} summarizes the results. Using RL-TNCO to optimize the last $k=100$, $1000$, or $2000$ steps after a greedy path prefix reduces the total flop count by factors of $\times1.09$, $\times 1.9$, and $\times3.9$, respectively. CTG-greedy and CTG-kahypar ran with a time limit of 6 hours.

\setlength{\tabcolsep}{5pt}
\begin{table}
\centering
  \small
  \caption{Comparison of cost and runtime for Max-cut TNs.}
    \vspace{5pt}
    \resizebox{0.5\linewidth}{!}{%
    \begin{tabular}{lrr}
    \toprule
    Method      & contraction cost & {time} \\
    {~}         & $\times 10^{12}$ flops & {~} \\
    \midrule
    OE-Greedy   &   36.45 & 5.1 sec \\ 
    CTG-Kahypar &  144 & 6 hours\\
    CTG-Greedy  & 2540 & 6 hours\\
    \midrule
    RL-TNCO (last 100 steps)  & 33.40 & {20 min} \\
    RL-TNCO (last 1000 steps) & 19.05 & {1.5 hrs}\\
    RL-TNCO (last 2000 steps) & \textbf{9.29} & {5.5 hrs} \\
    \bottomrule
    \end{tabular}%
    }
  \label{tab:large_scale}%
\end{table}%

\subsection{Ablation} 

To understand the contribution of components of our model, we performed an ablation study for the four components discussed in Section \ref{sec:appproach}. Table \ref{tab-ablation} presents the results of the ablation study. The ``Full" model is often the best configuration. Each component of RL-TNCO seems to improve the final result. Integrating existing solver information seems to have the most pronounced effect.  

\begin{figure}
    \centering 
    \includegraphics[scale=0.45]{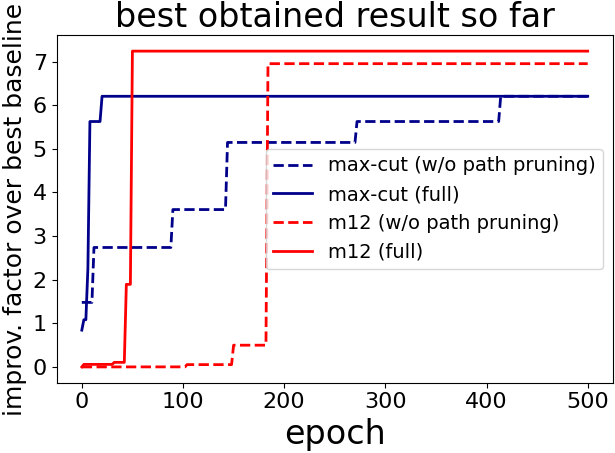}
    \caption{In a single-network setup we keep the best result obtained during training. This plot show the best improvement factor (best baseline FLOPS/FLOPS) as a function of the training epoch. Path pruning allows RL-TNCO to achieve a good result faster.}
    \label{fig:TN}
\end{figure}

Path pruning trades critic accuracy for train time. Fig \ref{fig:TN} shows that a good result is achieved faster with path pruning.

\begin{table}
    \centering
    \setlength{\tabcolsep}{3pt}
    \small
    \begin{tabular}{l|ccccc}
    &  & w/o  & w/o   & w/o  & w/o   \\
     & Full & robust & solver  & optimistic & path  \\
     &  & features & info & buffer & pruning \\
    \midrule
    random TN & $\boldsymbol{\times44.3}$ & $\times12.6$ & $\times4.5$ & $\times6.1$ & N/A \\
    150 nodes & & & & & \\ \midrule
    max-cut,  & $\times6.2$ & $\times3.6$ & $\times0.7$ & $\times6.2$ & $\times6.2$ \\ 
    last 100  & & & & & \\ \midrule
    m12 & $\boldsymbol{\times7.2}$ & $\times3.3$ & $\times10^{-3}$ & $\times6.0$ & $\times6.9$ \\
    \bottomrule
    \end{tabular}
    \caption{Improvement factor (best baseline FLOPS/RL-TNCO FLOPS) over best baseline for ablated models on various TNs. Path pruning during training is applicable in single network mode only. The max-cut results are for the suffix TN, excluding the shared prefix TN.}
    \label{tab-ablation}
\end{table}

\subsection{Runtime vs performance trade-off: The Sycamore M20 circuit} 
In this subsection, we present a detailed performance comparison of RL-TNCO on the largest Sycamore circuit, Sycamore M20. A recent experiment on this circuit is claimed to prove quantum supremacy \cite{arute2019quantum}.

\paragraph{Optimizing time-constrained performance.}  When learning on a single network, one can improve performance by either using multiple seeds, resulting in different trained models or increasing the training time \footnote{In setup 2: Learning and generalizing to new networks, it is also possible to increase the number of sampled trajectories.}. Likewise, other solvers may benefit from using multiple seeds, or extending the evaluation time.  Our goal in this section is to investigate the performance gain when the the time budget is extended. This would allow us to understand what the optimal configuration is for a given time budget and in which configuration extending the solver time limit will result in decreasing marginal utility. 

 Both RL-TNCO and Cotengra-Kahypar are not deterministic, and we find that results tend to vary considerably across different initializations (different seeds). It therefore makes sense to run multiple instances with different seeds and select the best result. We are interested in estimating the result of each baseline with time limit $\tau$ and $k$ seeds. Since this is an optimization score, the baseline score is the best score over the $k$ seeds. The time limits ranged from 30 minutes to 3 hours, and we used up to 10 different seeds for each baseline.



\paragraph{Analysis.} We used the bootstrap analysis method \cite{singh2008bootstrap} to evaluate the median result. We run this baseline with 10 seeds for each time limit  $\tau$ and obtained $10$ results. We sampled with replacement $n\cdot k$ results, where $k=2..10$, and obtained $n$ tuples of size $k$. For each tuple $x_i(\tau,k)$ we took the best result, yielding $n$ values $z_i(\tau,k)$. We report the median $m(\tau,k)$ over these values $\{z_i(\tau,k)|i=1..n\}$ for Contegra Kahypar in Table \ref{tab:contegra_m20_detailed} and for RL-TNCO in Table \ref{tab:rl_m20_detailed}. In this graph, the cost of calculating the path order is negligible compared to the contraction time , and the gain from reducing contraction cost is by far higher than longer running times (Table \ref{tab:m20_analysis}).

\begin{table} 
\begin{centering}
\begin{tabular}{|cc|ccccccccc|}
\hline 
\multirow{2}{*}{} & \multirow{2}{*}{} & \multicolumn{9}{c|}{Number of seeds} \tabularnewline
\cline{3-11} \cline{4-11} \cline{5-11} \cline{6-11} \cline{7-11} \cline{8-11} \cline{9-11} \cline{10-11} \cline{11-11} 
 &  & 2 & 3 & 4 & 5 & 6 & 7 & 8 & 9 & 10\tabularnewline
\hline 
\hline 
\multirow{4}{*}{\rotatebox[origin=c]{90}{Limit (hrs)}} & 0.5 & 7.54 & 7.53 & 7.475 & 7.42 & 7.42 & 7.42 & 7.14 & 7.14 & 7.14 \tabularnewline
 & 1 & 7.53 & 7.53 & 7.48 & 7.48 & 7.48 & 7.48 & 6.01 & 6.01 & 6.01\tabularnewline
 & 2 & 6.7 & 6.45 & 6.43 & 6.43 & 6.43 & 6.43 & 5.83 & 5.83 & 5.83\tabularnewline
 & 3 & 6.7 & 6.45 & 6.43 & 6.43 & 6.43 & 6.43 & 5.83 & 5.83 & 5.83\tabularnewline 
\hline 
\end{tabular}\caption{The median contraction cost (in $10^{18}$ flops) for Sycamore M20 circuit using Contegra-Kayhpar. The estimation is based on the booststrap method. For the 3 hours time limit runs, we present the best result out of the 2 hour run and 3 hour run} \label{tab:contegra_m20_detailed}
\par\end{centering}
\end{table}

\begin{table}
\begin{centering}
\begin{tabular}{|cc|ccccccccc|}
\hline 
\multirow{2}{*}{} & \multirow{2}{*}{} & \multicolumn{9}{c|}{Number of seeds} \tabularnewline
\cline{3-11} \cline{4-11} \cline{5-11} \cline{6-11} \cline{7-11} \cline{8-11} \cline{9-11} \cline{10-11} \cline{11-11} 
 &  & 2 & 3 & 4 & 5 & 6 & 7 & 8 & 9 & 10\tabularnewline
\hline 
\hline 
\multirow{4}{*}{\rotatebox[origin=c]{90}{Limit (hrs)}} & 0.5 & \cellcolor{red!25}5864 & \cellcolor{red!25}93.1 & \cellcolor{red!25}13.13 & \cellcolor{red!25}13.13 & \cellcolor{red!25}13.13 & \cellcolor{red!25} 13.13 & \cellcolor{red!25}9.79 & \cellcolor{red!25}9.79 & \cellcolor{red!25}9.79\tabularnewline
 & 1 & \cellcolor{red!25}5864  & \cellcolor{red!25}93.1 & \cellcolor{red!25}9.79 & \cellcolor{red!25}9.79 & \cellcolor{red!25}9.79 & \cellcolor{red!25}9.79 & \cellcolor{red!25}6.81 & \cellcolor{red!25}6.81 & \cellcolor{red!25}6.81\tabularnewline
 & 2&  \cellcolor{red!25}127.5 & \cellcolor{red!25}93.1 & \cellcolor{red!25}7.38 & \cellcolor{red!25}7.38 & \cellcolor{red!25}7.38 & \cellcolor{red!25}7.38 & \cellcolor{green!25}5.25 & \cellcolor{green!25}5.25 & \cellcolor{green!25}5.25 \tabularnewline
 & 3 & \cellcolor{red!25}37.63 & \cellcolor{red!25}6.68 & \cellcolor{green!25}5.25 & \cellcolor{green!25}5.25 & \cellcolor{green!25}5.25 & \cellcolor{green!25}5.25 & \cellcolor{green!25}3.49 & \cellcolor{green!25}3.49 & \cellcolor{green!25}3.49\tabularnewline
\hline 
\end{tabular}\caption{The median contraction cost (in $10^{18}$ flops) for Sycamore M20 circuit using RL-TNCO. The estimation is based on the bootstrap method. Configuration which achieves superior performance over the corresponding setup using Cotengra-Kayhpar are highlighted in green, otherwise the cell is highlighted in red. The optimal result can be achieved with 8 seeds trained for three hours. \label{tab:rl_m20_detailed}}
\par\end{centering}
\end{table}

\section{Conclusions}
In this paper, we introduce the TNCO problem to the machine-learning community and propose the first ML-based method for solving it. We formulated the TNCO problem as an MDP, discussed the main challenges that arise from this formulation, and finally devised a novel approach based on RL and GNNs. The method outperforms the state of the art on several tensor network types. 

In a recent work, \cite{huang2021efficient} showed that performing a quantum simulation of recent large scale circuits, including Sycamore circuits, requires roughly 20 days on a cluster whose size is comparable to Summit. We show here that using RL to optimize contraction order can further reduce computing time by a factor of 2x-4x, minimizing highly expensive and scarce super-computing resources, reducing the amount of energy consumption of simulating just a single circuit by GWhs, amounting to powering hundreds of household per year, saving tens of thousands of dollars, and consequently accelerating and advancing research in QC.

\bibliography{arxiv}
\bibliographystyle{apalike}

\newpage
\appendix
\newpage
\appendix
\onecolumn
\section{Additional Experimental details}
\subsection{Hyperparameters}
Table \ref{tab:hyper} lists the hyperparameter values that were used for our trainable modules. Both the actor and the critic shared the same hyperparmeters.
Additioanlly, we scaled the rewards (FLOPS) by $10^{14}$ for the syntethic networks and the Sycamore circuit $m=10,12,14$, while for the max-cut and sycamore and M20 circuits we used $10^{19}$ as our scaling factor.

\begin{table}[h]
\begin{centering}
\begin{tabular}{|c|c|c|c|}
\hline 
Category & Parameter & synthetic networks & real-world networks \\
\hline 
\hline 
\hline 
Random TNs  & mean connectivity & $3$ & - \tabularnewline
\hline 
Random TNs  & tensor extent probability & $U[2,6]$ & -\tabularnewline
\hline 
GNN parameters & GNN layers  & $3$ & $4$ \\
\hline 
GNN parameter & \# intermediate layer features & $128$  & $256$ \\
\hline 
GNN parameter & Aggregation method & mean & mean\tabularnewline
\hline 
GNN parameter & Use Bias & False & False \tabularnewline
\hline 
RL & roll-out steps & $4192$ & $2096$ \\
\hline 
RL & Training samples & $1M$ & $3M$ \\
\hline 
RL & batch size & $512$ & $256$ \tabularnewline
\hline 
RL  & Value weight & $0.1$ & $0.001$\tabularnewline
\hline 
RL & Score Normalization, $\epsilon$ & $10^{-2}$ & $10^{-2}$\tabularnewline
\hline 
RL   & \# paths during inference per inference step& 50, 1000 & $160$\tabularnewline
\hline 
RL   & Inference step frequency & - & $40$ \\
\hline 
RL & PPO training step per roll-out collection & $8$ & $8$\tabularnewline
\hline 
\hline 
 RL & Learning rate & $3\cdot10^{-2}$ & $3\cdot10^{-2}$\tabularnewline
\hline 
RL  & Entropy weight & $0$ & $0$\tabularnewline
 \hline

RL  & Optimistic buffer size & $4096$ & $4096$\tabularnewline
  \hline 
\end{tabular}
\par\end{centering}
\caption{Hyperparameters table}
\label{tab:hyper}
\end{table}




\subsection{Wall-time Comparison} 
Table \ref{tab:wallclock} compares the total cost of finding the optimal contraction path and the actual time of executing the contraction for different algorithms and networks. Recall that for QC applications, each network has to be contracted multiple times to get a high-fidelity estimate of the circuit output \cite{peng2021graph}. Here, we assume that we contract each network $\ell=10^6$ times \cite{pan2021solving}. The results indicate that RL-TNCO is 90\%-40\% faster than the best baseline considered.
\setlength{\tabcolsep}{2pt}
\begin{table}[htbp]
    \centering
    \small
    \begin{adjustbox}{max width=\textwidth}
    \begin{tabular}{l|rrrr | rrrr|c|}
    & \multicolumn{4}{c}{RL-TNCO}   & \multicolumn{4}{c}{Best baseline} &  \\
    \midrule
    & \multicolumn{1}{c}{path } 
    & \multicolumn{1}{c}{contract.} & 
    \multicolumn{1}{c}{single} & 
    \multicolumn{1}{c}{total} & 
    \multicolumn{1}{c}{path} & 
    \multicolumn{1}{c}{contract.} & 
    \multicolumn{1}{c}{single } & 
    \multicolumn{1}{c}{total } & 
    \multicolumn{1}{c}{RL-TNCO} \\
    & \multicolumn{1}{c}{optim`} & 
    \multicolumn{1}{c}{cost} & 
    \multicolumn{1}{c}{contract.} & 
    \multicolumn{1}{c}{} & 
    \multicolumn{1}{c}{optim`} & 
    \multicolumn{1}{c}{cost } & 
    \multicolumn{1}{c}{contract.} & 
    \multicolumn{1}{c}{} & 
    \multicolumn{1}{c}{to} \\
    network 
    & \multicolumn{1}{c}{(sec)} & 
    \multicolumn{1}{c}{(flops)} & 
    \multicolumn{1}{c}{(sec)} & 
    \multicolumn{1}{c}{(sec)} & 
    \multicolumn{1}{c}{(sec)} & 
    \multicolumn{1}{c}{(flops)} & 
    \multicolumn{1}{c}{(sec)} & 
    \multicolumn{1}{c}{(sec)} & 
    \multicolumn{1}{c}{Best baseline} \\
    \midrule
    Random 100 & 1.6E+02 & 5.5E+12 & 2.7E-01 & 2.7E+05 & 1.0E+04 & 5.4E+13 & 2.7E+00 & 2.7E+06 & 0.10 \\
    Sycamore m=14 & 4.3E+04 & 2.6E+14 & 1.3E+01 & 1.3E+07 & 1.0E+04 & 5.0E+14 & 2.5E+01 & 2.5E+07 & 0.52 \\
    Sycamore m=20 & 4.3E+04 & 3.5E+18 & 1.7E+05 & 1.7E+11 & 1.0E+04 & 5.8E+18 & 2.9E+05 & 2.9E+11 & 0.60 \\
    MaxCut & 8.6E+04 & 9.2E+12 & 4.6E-01 & 5.5E+05 & 2.1E+04 & 3.1E+13 & 1.5E+00 & 1.5E+06 & 0.35 \\
    \midrule
    \end{tabular}%
    \end{adjustbox}
    \caption{Wall time Comparison of path ordering and actual contraction time. Time is measured in seconds, cost in flops. Conversion is based on DGXA100 specifications.}
  \label{tab:wallclock}%
\end{table}%

\end{document}